\documentclass{aa}
\usepackage[varg]{txfonts}
\usepackage{color}
\usepackage{ulem}

\begin{document}

\title{On the lifetime of merger features of equal-mass disk mergers}

\author{Inchan Ji\inst{1}, S\'ebastien Peirani\inst{2}, and Sukyoung K. Yi\inst{1}}

\institute{Department of Astronomy and Yonsei University 
Observatory, Yonsei University, Seoul 120-749, Republic of Korea\\
\email{yi@yonsei.ac.kr} 
\and
Institut d'Astrophysique de Paris (UMR 7095: CNRS \& UPMC), 
98 bis Bd Arago, 75014 Paris, France\\}

\abstract
{Detecting post-merger features of merger remnants is highly dependent 
on the depth of observation images. 
However, it has been poorly discussed how long the post-merger 
features are visible under different observational conditions.
}
{
We investigate a merger-feature time useful for understanding 
the morphological transformation of galaxy mergers 
via numerical simulations. 
}
{We use N-body/hydrodynamic simulations, 
including gas cooling, star formation, and supernova feedback.  
We run a set of simulations
with various initial orbital configurations and with progenitor 
galaxies having different morphological properties mainly 
for equal-mass mergers.
As reference models, we ran additional simulations for 
non-equal mass mergers and mergers in a large halo potential.
Mock images using the SDSS $r$ band are synthesized  
to estimate a merger-feature 
times and compare it between the merger simulations.
} 
{
The mock images 
suggest that the post-merger features involve a small fraction of stars, 
and the merger-feature time depends on galaxy interactions. 
In an isolated environment, the merger-feature time is, 
on average, $\sim$ 2 times the final coalescence time for a 
shallow surface bright limit of 25 mag arcsec$^{-2}$. 
For a deeper surface brightness limit of 28 mag arcsec$^{-2}$, 
however, the merger-feature time is a factor of two 
longer, which is why the detection of post-merger features 
using shallow surveys has been difficult. 
Tidal force of a cluster potential is effective in stripping  
post-merger features out and reduces the merger-feature time.
}
{%
}
\date{Accepted 2014 3 May}
 
\maketitle

\section{Introduction}

Since the seminal work of \citet{TT72} illustrating the peculiar features 
of nearby galaxies, galaxy merger has been considered a fundamental 
phenomenon in understanding galaxy formation. The Lambda Cold 
Dark Matter ($\Lambda$CDM) model predicts the hierarchical galaxy formation 
which suggests that present-day galaxies are formed from successive 
accretions and mergers of smaller entities \citep{White78}. In this context, large 
early-type galaxies are believed to have experienced mergers between 
disk galaxies in the past \citep{Toomre77}.

Theoretical perspectives of galaxy 
interactions help us understand galaxy formation and evolution. 
In particular, star formation 
histories and the dynamical properties of galaxy mergers 
have been investigated through numerical simulations 
\citep [e.g.][]{Gerhard81,Barnes91,Hernquist92,Naab03,Cox06a,
Cox06b,DiMatteo07, Peirani10}. 
The numerical simulations suggest that the properties of merger remnants, 
such as their isophotal shapes, surface density profiles, 
and the ratio between local rotation to velocity dispersion (v/$\sigma$), are
 consistent with those of observed early-type galaxies. In addition,
 dissipative mergers can account for the central concentration of starbursts
 between merger remnants \citep{Barnes91,Mihos94,Springel00}.

However, the gap between observation and 
theoretical study is still considerable. 
The properties of galaxies beyond their internal structures, 
e.g., magnitudes and colors, have rarely been studied 
primarily due to the difficulty of modeling 
dust effects in the interstellar medium of galaxies. 
Various methods have recently been developed to overcome this issue.

One method is the use of Monte Carlo radiative transfers 
assuming idealized galaxy models or using distributions 
of gas and stars from numerical simulations \citep{Silva98,Jonsson06}. 
Although this method provides detailed and extensive panchromatic data 
for galaxies, it requires high computational costs.
The other method is to directly assign spectral 
data to a single stellar population using population synthesis models. 
This method is effective because of its simplicity and applicability. 
Several studies have examined the observable properties of merger remnants 
by assuming the total amount of dust extinction from empirical data 
\citep{Kaviraj09,Gabor11}. Therefore, their approaches are well-suited 
to studying the integrated magnitudes and colors of each galaxy, 
but not to studying regional differences. 

The data provided by large surveys and improvements 
in numerical modeling give us an opportunity to study all physical 
processes that occur during the interactions of galaxies. 
As an illustration, large surveys such as the CfA2 Redshift Survey, 
the Two-Degree Field (2dF) Galaxy Redshift Survey, the Deep Extragalactic 
Evolutionary Probe (DEEP), and the Sloan Digital Sky Survey (SDSS) 
have collected detailed galaxy properties, while hydrodynamic 
simulations and population synthesis models have been employed  
to produce the observable properties of merger remnants. 
Recent studies have tried to compare the photometric quantities derived 
from numerical simulations to data from large surveys 
\citep{Springel05d,Gabor12}.

Most massive galaxies are expected to undergo a series of merger events 
\citep{Stewart08}. It is then necessary to constrain timescales such as 
first passage, maximum separation, and final coalescence when 
we predict the physical properties of galaxies. 
Observations help identifying  the timescales using angular separation, 
line-of-sight radial velocity, and correlation functions 
\citep{Patton00,Barton00,Lin04,dePropris05,Bell06,Li08}. 
Recent studies have shown that the timescales before the final 
coalescence can be 
formulated analytically based on numerical simulations 
\citep[][]{Boylan-Kolchin08,Jiang08}. However, only a few studies 
have measured timescales beyond the final coalescence. 
One such timescale is the merger-feature time, the moment when 
faint and disturbed 
features of merger remnants (post-merger features) disappear. \citet{Lotz08} 
assumed that the merger-feature time (``post-merger'' stage in the literature) 
ends 1 Gyr after the final coalescence. 
However, the merger-feature time obviously depends on 
observational conditions such as surface brightness limits.

\begin{figure}
\centering
\includegraphics[width=0.45\textwidth]{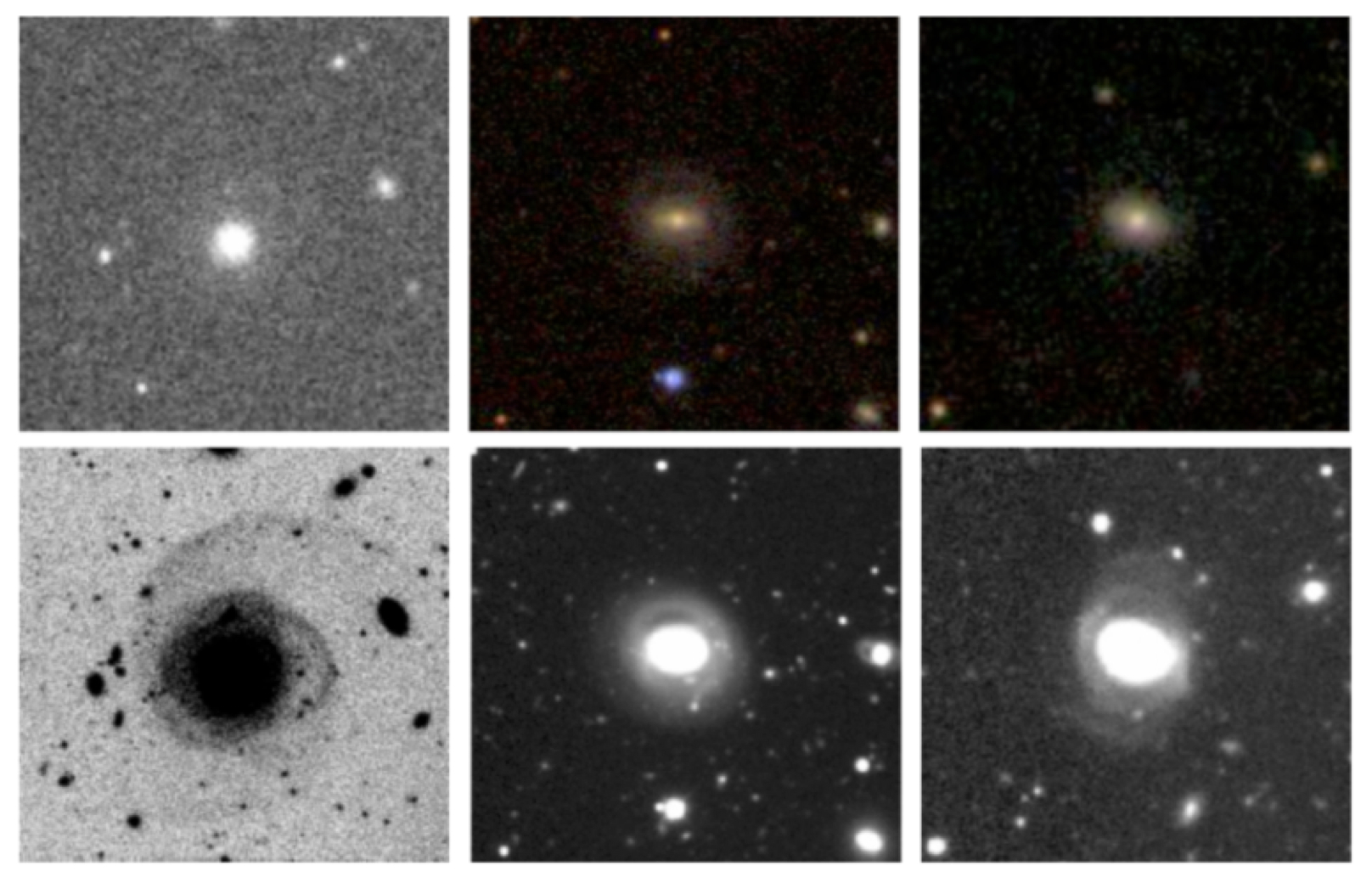}
\caption{The SDSS images (top) and CTIO $r'$ band images (bottom) 
of galaxies in Abell 2670. The images are courtesy of Yun-Kyeong Sheen. 
The galaxies in each column are the same object. The surface bright limits 
in the SDSS and CTIO observations are $\mu_{\rm limit} \sim$ 25 and 28 
mag arcsec$^{-2}$, respectively. One can see that the deep images reveal 
faint tidal features around the galaxies.}
\label{merger_feature}
\end{figure}

Figure~\ref{merger_feature} shows three early-type galaxies 
in Abell 2670 observed from the SDSS ($\mu_{\rm limit} \sim$ 
25 mag arcsec$^{-2}$)
and from the CTIO $r'$ band ($\mu_{\rm limit} =$ 28 mag arcsec$^{-2}$)
by \citet{Sheen12}. The images from the SDSS show only the central regions 
of the galaxies and suggest elliptical-like galaxies. 
However, the deeper images from the CTIO observations 
clearly show faint structures around the galaxies and
thus characterize more accurately the dynamical and relaxation 
state of the galaxies. Therefore, this suggests that the merger-feature time 
could be estimated more accurately with the deeper images.

In this work, our aim is to study the merger-feature time  
after galaxy merger while considering both reasonable dynamical 
evolutions of galaxy mergers and dust attenuation. We produce 
a set of merger events using N-body/hydrodynamic simulations from 
which we extract the dust-attenuated spectra of galaxies using both 
\citet{BC03} and \citet{Calzetti00}. 
This modeling allows us to construct mock images of galaxies mimicking 
observational environments to be compared with observational data.

This paper is organized as follows: the numerical modeling is summarized in 
Section \ref{section_numerical}, while the methodology used to extract 
observable properties is described in Section \ref{analysis}. 
In Section \ref{results}, we 
compare the merger-feature times of merger remnants. Lastly, we discuss 
the results in Section \ref{discussion}.

\section{Numerical modeling }
\label{section_numerical}
\subsection{Galaxy modeling and simulations}

The numerical methodology used in this paper is described in detail in 
\citet{Peirani09} to which we refer the reader for more information. 
For the sake of clarity, we summarize the major steps below.

Our study is focused on 1:1 mass-ratio 
mergers where the final coalescence time is the shortest in the large 
parameter space \citep[][for example]{Boylan-Kolchin08}. This fact is 
also useful for measuring various timescales as we will discuss below. 
Galaxies are created following  the prescription described by 
\citet{Springel05b}. We use a compound 
galaxy model consisting of a spherical dark matter halo, 
a rotationally-supported disk (of gas and star),  
and a bulge, with independent parameters describing each of the structural 
components. The distribution of the dark matter halo follows a Hernquist profile 
\citep{Hernquist90}, with a concentration parameter of $C = 14$ in both 
galaxies following  \citet{Dolag04}. We also assume a thin disk structure 
of gas and stars with an exponential surface density. The scale length of 
disk $R_{\rm d}$ is determined following  \citet{Graham08}, and the bulge 
distribution is assumed to be spherical 
while also following a Hernquist profile. We adopt a conventionally 
determined disk scale height $z_0$, and a bulge scale length $b$ 
as $b = z_0 = 0.2~R_{\rm d}$. The baryon fraction derived from 
\citet{Komatsu11} is used. The gas fraction of our model galaxies is 
consistent with the 
estimation in \citet{BM98} and \citet{Kannappan04}. The maximum speed 
of the rotation curve is about 150 km/s as suggested by the baryonic 
Tully-Fisher relation \citep{McGaugh05}. We limit our study to Sa and Sb 
model galaxies.
To consider unequal-mass mergers, we constructed two satellite galaxies 
by changing the total mass of the Sb galaxy while fixing the other 
parameters. They are one-third and one-sixth of the Sb galaxy in total 
mass and are labelled as  
${\rm Sb_{3}}$, ${\rm Sb_{6}}$, and ${\rm Sb_{10}}$, respectively.
Table \ref{initgal} summarizes the parameters of each galaxy.

\begin{table*}
\begin{center}
\caption{Initial parameters of model galaxies}

\begin{tabular}{lccccc} 
\hline
	 {} & Sa & Sb & ${\rm Sb}_3$ 	&${\rm Sb}_6$  &${\rm Sb}_{10}$\\
\hline
  
	Total mass, $M_{\rm vir}~({\rm 10^{10} M}_{\odot})$ &17.0	&17.0 &5.7 &2.8 & 1.7\\
	DM mass, $M_{\rm DM}~({\rm 10^{10} M}_{\odot})$ &14.1 &14.1	 &4.7 &2.4 &1.4\\ 
	Stellar disk mass, $M_{\rm d}~({\rm 10^{10} M}_{\odot})$ &1.61 &1.81 &0.60 &0.30 &0.18\\
	Stellar bulge mass, $M_{\rm b}~({\rm 10^{9} M}_{\odot})$ &10.74 &4.52 &1.51 &0.75	&0.45\\
	Gas mass, $M_{\rm gas}~({\rm 10^{9} M}_{\odot})$ &1.41 &5.65	&1.88 &0.94 &0.57\\
	B/T ratio	&0.4	&0.2	&0.2 &0.2 & 0.2\\
	Spin parameter, $\lambda$ &0.05 &0.05 &0.05	&0.05 & 0.05\\		
	Disk scale length, $R_{\rm d}~{\rm (kpc)}$ 	&2.58 &3.29 &2.28 &1.81 &1.52\\
\hline
\end{tabular}
\label{initgal}
\end{center}
\end{table*}

The simulations were performed using the publicly-available code 
GADGET2 \citep{Springel05a} with added prescriptions for gas cooling, 
star formation, and feedback from Type Ia and II supernovae (SNe). 
Note that gas particles with $T< 2\times 10^4~{\rm K}$, number density 
$n > 0.1~{\rm cm}^{-3}$, overdensity $\Delta \rho_{gas}> 100$, and 
${\bf \nabla . \upsilon} <0$ form stars according to 
the standard star formation prescription: 
$d\rho_*/dt = c_* \rho_{\rm gas}/t_{\rm dyn}$, where $\rho_*$ refers to 
the stellar density, $t_{\rm dyn}$ is the local dynamical timescale 
of gas, and $c_*$ is the star formation 
efficiency. More than a million particles were used for each run. 
Consequently, the particle masses are $m_{\rm DM} = 4 \times 
10^{5}~{\rm M}_{\odot}$, $m_{\rm disk} = 
m_{\rm gas} = 5 \times 10^{4}~{\rm M}_{\odot}$, and $m_{\rm bulge} = 
1.5 \times 10^{5} ~{\rm M}_{\odot}$ for dark matter, disk, gas, and bulge, 
respectively. The gravitational softening lengths are $\epsilon_{\rm DM}$ = 
0.1~kpc, $\epsilon_{\rm disk} = \epsilon_{\rm gas}$ = 0.2~kpc, and 
$\epsilon_{\rm bulge}$ = 0.1~kpc.

\begin{table}
\caption{Initial orbital configurations}
\begin{tabular}{lcccc}

\hline
Simulation$^{\mathrm{a}}$ & 
Pair$^{\mathrm{b}}$ &
$e^{\mathrm{{d}}}$ &
$R_{\rm peri}^{\mathrm{d}}$&
$i_{\rm host}^{\mathrm{e}}$\\
\hline

	SbSb0p  		& Sb-Sb		&1		&5				&$0$	\\	
	SbSb45p  	& Sb-Sb		&1		&5				&$45$	\\
	SbSb90p  	& Sb-Sb		&1		&5				&$90$	\\
	SbSb135p  	& Sb-Sb		&1		&5				&$135$	\\
	SbSb180p  	& Sb-Sb		&1		&5				&$180$	\\

	SaSb45p  	& Sa-Sb		&1		&5				&$45$	\\
	SaSb90p  	& Sa-Sb		&1		&5				&$90$	\\
	SaSb135p  	& Sa-Sb		&1		&5				&$135$	\\

	SbSb45r 		& Sb-Sb		&1		&0				&$45$	\\
	SbSb45p+	& Sb-Sb		&1		&10				&$45$	\\
	
	SbSb45e 		& Sb-Sb		&0.95	&5				&$45$	\\
	SbSb45h		& Sb-Sb		&1.05	&5				&$45$	\\
	
	${\rm SbSb_{3}45h}$		& Sb-${\rm Sb_3}$		&1.05	&5				&$45$	\\
	${\rm SbSb_{6}45h}$		& Sb-${\rm Sb_6}$		&1.05	&5				&$45$	\\	
	SbSb45hC	& Sb-Sb		&0.95$^{\mathrm{f}}$	&472$^{\mathrm{f}}$				&-	\\
\hline	

\label{fiducial}
\end{tabular}
\begin{list}{}{}
\item[$^{\mathrm{a}}$] Types of galaxies, initial encounter angle, 
orbital shape ("p" for parabolic, "r" for radial, "e" for elliptical, 
"h" for hyperbolic, and "+" for larger pericentric distance), 
and additional cluster potential ("C" for cluster potential)
are marked in the simulation name.
\item[$^{\mathrm{b}}$] Host-companion pair.
\item[$^{\mathrm{c}}$] Eccentricity of orbit.
\item[$^{\mathrm{d}}$] Pericentric distance (kpc).
\item[$^{\mathrm{e}}$] Inclination of host galaxy ($^{\circ}$).
\item[$^{\mathrm{f}}$] They are the orbital parameters 
with respect to a cluster potential.
\end{list}
\end{table}

\begin{figure}
\centering
\includegraphics[width=0.5\textwidth]{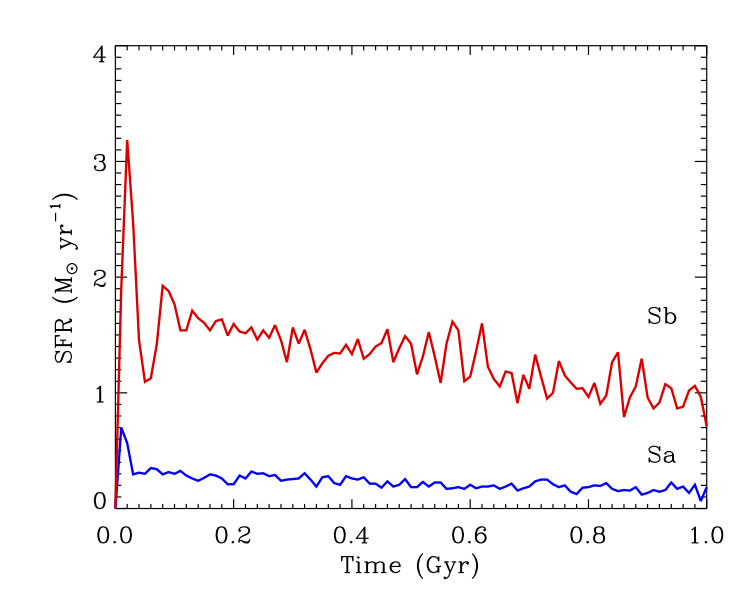}
\caption{Star formation rate for isolated Sa and Sb models at intervals 
of 10 Myr to 1 Gyr. 
The Sb model with a higher gas content shows star formation rates 
roughly an order of magnitude higher than the Sa model. 
Initial bursts of new stars in both models are contributed from  
the gas compression of gas into density waves 
and the absence of supernova feedback from young stellar populations.}
\label{sfr_consti}
\end{figure}

The stability of our model galaxies is measured by investigating 
the star formation rate (SFR) of each galaxy in isolation as shown 
in Figure~\ref{sfr_consti}. We found SFRs of, on average, 
$\sim 0.22$ M$_{\odot}$yr$^{-1}$ and $\sim 1.3~{\rm M}_{\odot}~$yr$^{-1}$ 
over 1 Gyr for the Sa and Sb models, respectively. Note that 
the SFRs are consistent with the observational sample of \citet{James04}.

The difference between the SFR amplitudes of the two model galaxies 
is due to the difference between the initial masses of gas in the disks 
(see Table~\ref{initgal}). The gradual decay of SFR is as expected since 
the available gas is progressively converted into stars. External accretion of gas 
could be considered to reproduce the constant average SFRs 
found in observations of spiral galaxies in the middle of the Hubble sequence 
\citep{Kennicutt83,Kennicutt94} but is not included in our study to simplify 
our simulation.

As shown in Figure~\ref{sfr_consti}, initial bursts of new stars occur as a result 
of gas compression into density waves early in the simulations. 
However, the initial bursts are suppressed after supernova feedbacks are 
exerted on gas \citep[see][]{DiMatteo07}. Each model galaxy is quite stable 
as no spurious effects are seen in the evolution of the SFRs 
due to specific choices of gravitational smoothing or the 
phase of the initial condensation of gas.

\begin{figure}
\centering
\includegraphics[width=0.45\textwidth]{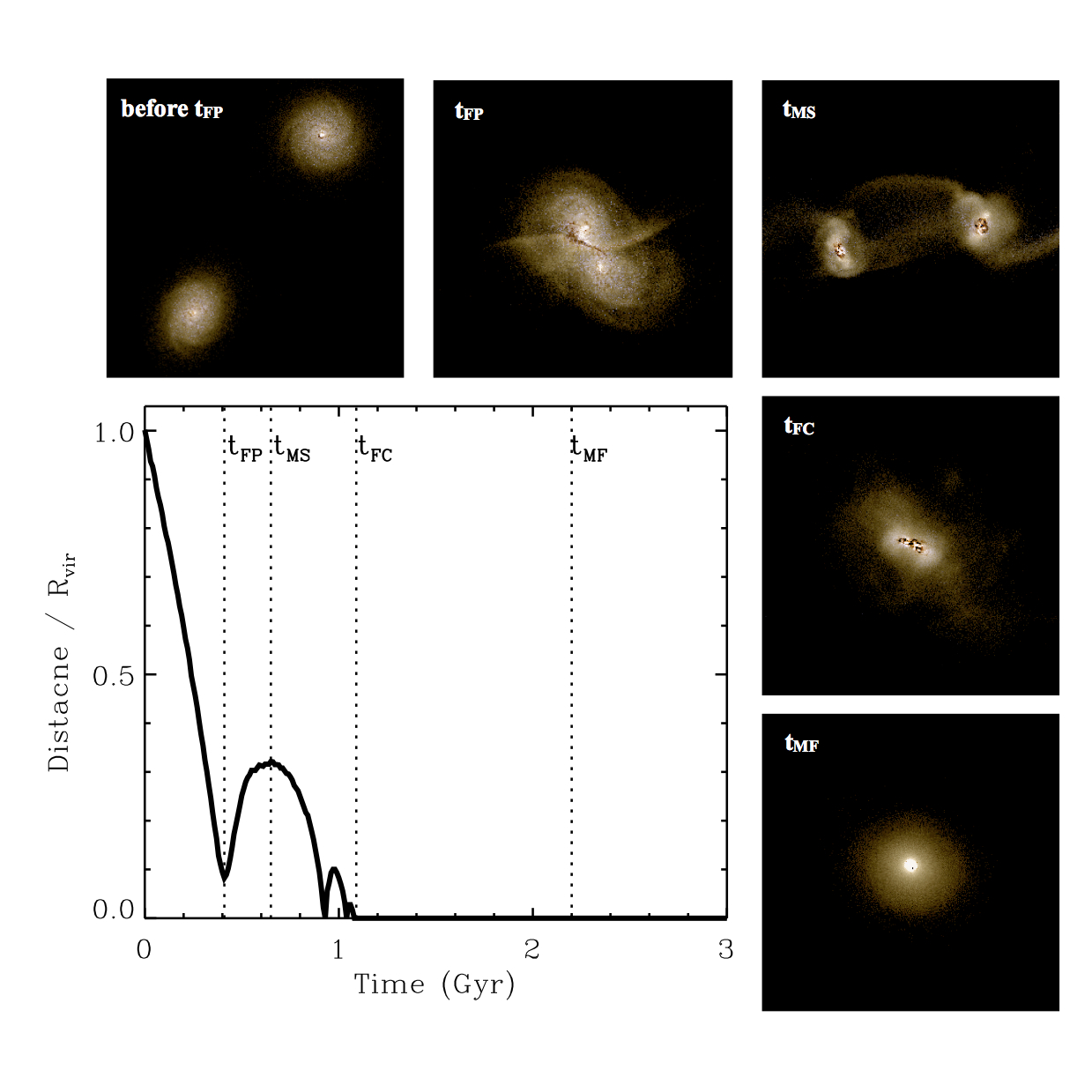}
\caption{Overview of merger timescales: 
first passage time ($t_{\rm FP}$) when a companion or satellite 
galaxy first passes the closest distance, 
maximum separation time ($t_{\rm MS}$) after the first passage,  
final coalescence time ($t_{\rm FC}$) when two galaxies finally merge, and 
merger-feature time ($t_{\rm MF}$) 
when post-merger features disappear given a surface brightness limit. 
}
\label{merger_step}
\end{figure}

\subsection{Definition of Merger Timescales}
\label{time_scales}

Figure \ref{merger_step} shows characteristic timescales of a merger 
event. Each merger event is characterized by several successive and 
distinct phases. We identify several critical timescales that show distinctive 
features during the merger: the first passage time 
($t_{\rm FP}$) when 
the companion galaxy first passes the closest distance, 
the maximum separation time ($t_{\rm MS}$) after 
the first passage time, 
and the final coalescence time ($t_{\rm FC}$) when 
the two galaxies finally merge. In addition, we adopt the merger-feature time 
($t_{\rm MF}$) when the merger remnant hides
post-merger features 
viewed perpendicular to and parallel to the orbital plane. 
We consider two merger-feature times by visual 
inspection based on two surface brightness limits in the SDSS $r$ band: 
$\mu_{\rm r,limit}$ = 25 ($t_{\rm MF,25}$) and 
28 mag arcsec$^{-2}$ ($t_{\rm MF,28}$). 
The $\mu_{\rm r,limit}$ =  
25 mag arcsec$^{-2}$ is comparable to the observational limit of the SDSS. 
The characteristic times are listed in Table \ref{mu28_time}.

\subsection{Initial orbital configurations}

In all simulations, the initial separation between the host and the companion 
is 90 kpc, the virial radius of the host galaxy. The pericentric distances adopted 
here are consistent with those of dark matter haloes in cosmological 
simulations \citep[see][]{Khochfar06}. The definition of inclination with 
respect to the orbital plane is adopted from \citet{TT72}.

Table \ref{fiducial} summarizes all information related to  the specific orbital 
configuration of each merger. For the convenience of comparison between the 
simulations, we first constructed five fiducial mergers (top five mergers in 
Table \ref{fiducial}). The progenitor galaxies 
are Sb models. The eccentricity is e = 1, and the pericentric distance is 
$R_{\rm peri}$ = 5 kpc. The inclinations of the host galaxies are set to  
$i_{\rm host} = 0^{\circ}$ (prograde)$,~45^{\circ},~90^{\circ},~135^{\circ}$, and 
$180^{\circ}$ (retrograde).

Modifying the fiducial mergers, we constructed ten 
additional mergers. First, the host galaxy was altered from an Sb to an Sa 
galaxy, and the inclinations of the host galaxies are $i_{\rm host} = 
45^{\circ},~90^{\circ}$, and $135^{\circ}$. 
Second, the pericentric distance of the SbSb45p merger was changed so that  
$R_{\rm peri}$ = 0 and 10 kpc. Third, the orbital 
eccentricity of the SbSb45p merger was varied to consider elliptical 
and hyperbolic orbits as well: 
we chose e = 0.95 and 1.05. Forth, we changed the total mass of the 
companion galaxy in the SbSb45h merger so that mass ratios are 3:1,  
6:1, and 10:1. 
We chose the SbSb45h merger for comparison because its 
merger timescales are relatively shorter than those in other 
cases and so better suited for demonstrating the effects of merger 
conditions in question.

Lastly, we placed a merger remnant in a rigid dark matter potential 
following a Hernquist profile with a Virgo-like mass of ${M_{\rm vir}} 
\simeq 4.2 \times {\rm 10^{14}~M}_{\odot}$, 
a concentration parameter of  $C = 9$, and a scale length
of $a = 236$ kpc. We chose the merger remnant of the SbSb45h 
merger at the final coalescence time and let the merger remnant 
orbit in the cluster potential; the initial 
distance from the center is 1200 kpc which is the virial radius of the 
cluster halo potential, and the pericentric distance is twice 
the scale length of the potential.
The choice of the merger in a cluster halo environment is largely 
arbitrary, and we would only like to check out the environmental 
effect in rough senses. In all, we simulated and analyzed fifteen 
merger events considering different host-companion pairs, 
inclinations, and orbits. All equal-mass mergers are simulated 
until 8.5 Gyr corresponding to $\sim$ 8 
times the dynamical timescale at the 
virial radius of the host galaxy ($t_{\rm dyn}\sim$ 1.1 Gyr). 
Since unequal-mass mergers evolve more slowly than equal-mass 
mergers, they are run for 13 Gyr, instead.

\subsection{Star Formation}

\begin{figure}
\centering
\includegraphics[width=0.5\textwidth]{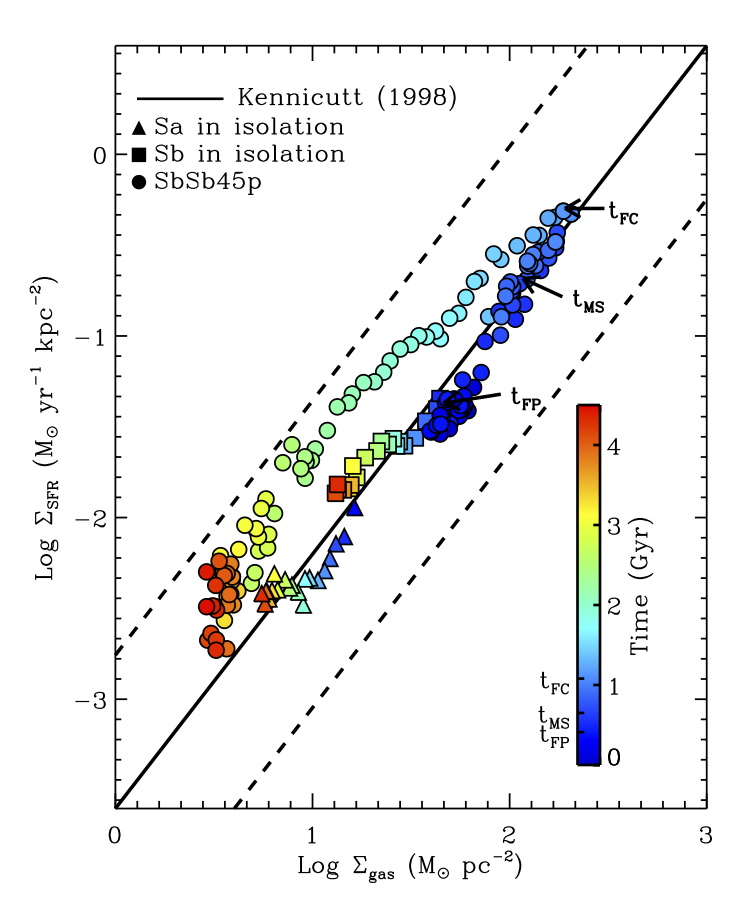}
\caption{Evolution of star formation surface density versus gas column 
density for the Sa model (filled triangle), Sb model (filled square), 
and SbSb45p merger (filled circle). Each symbol corresponds to the 
center of a galaxy at 5 and 20 Myr 
time intervals for each isolated progenitor (Sa and Sb) and the SbSb45p 
merger. For the SbSb45p merger, the merger timescales 
associated with star formation are labeled with arrows: 
i.e., the first passage time ($t_{\rm FP}$), 
the maximum separation time ($t_{\rm MS}$), 
and the final coalescence time ($t_{\rm FC}$).
The solid line is the best fit given by 
\citet{Kennicutt98}, and the dashed line represents 
a factor of 7 range which covers 
all observations in the same literature.}
\label{Kennicutt}
\end{figure}

As mentioned above, star formation in this study is intended to follow the 
empirical relation described in \citet{Kennicutt98}. Thus, to investigate 
the validity of our models, we also test the time evolution of the surface density 
of the gas and the star formation surface density averaged in the central region 
of each galaxy.

Figure~\ref{Kennicutt} shows the evolution of our model galaxies 
in isolation and one merger process (SbSb45p merger). 
Each symbol denotes the properties of a galaxy at an epoch given 
by the color key. In the isolated 
model galaxies, the surface density of gas and the star formation surface 
density decrease continuously starting from 
(log $\Sigma_{\rm gas}$, log $\Sigma_{\rm SFR}$) $\sim$ 
(1.2 ${\rm M_\odot~pc^{-2}}$, -2 ${\rm M_\odot~yr^{-1}~pc^{-2}}$) and
(log $\Sigma_{\rm gas}$, log $\Sigma_{\rm SFR}$) $\sim$ 
(1.7 ${\rm M_\odot~pc^{-2}}$, -1.3 ${\rm M_\odot~yr^{-1}~pc^{-2}}$) 
for the Sa and Sb models, respectively. Theses values closely track 
the empirical relation in \citet{Kennicutt98}. 

The SbSb45p merger, on the other hand, shows that the surface density of gas 
and the star formation surface density increase as the progenitor galaxies 
interact and merge with each other. During the merger-induced starburst, 
the gas surface density and the star formation surface density reach 
(log $\Sigma_{\rm gas}$, log $\Sigma_{\rm SFR}$) $\sim$ 
(2.3 ${\rm M_\odot~pc^{-2}}$, -0.2 ${\rm M_\odot~yr^{-1}~pc^{-2}}$)  
which are comparable to the ``starburst'' galaxies with 
(log $\Sigma_{\rm gas}$, log $\Sigma_{\rm SFR}$) $\gtrsim$ 
(2 ${\rm M_\odot~pc^{-2}}$, -1~${\rm M_\odot~yr^{-1}~pc^{-2}}$) 
in \citet{Kennicutt98}. 
Figure \ref{Kennicutt} also indicates that the SbSb45p merger evolves 
within the observational envelope. In addition, we confirm that there 
is good accordance between all merger simulations and the 
observational envelope. Therefore, we 
conclude that the model galaxies and the mergers discussed in this 
study closely follow the empirical star formation law, hence passing   
a sanity check regarding the sensibility of the star formation 
prescription in our models.

\section{Analysis}
\label{analysis}

\subsection{Modeling 2D Synthetic Images}
In this section, we describe our ray-tracing algorithm which allows the 
extraction of observable quantities from snapshots of merger simulations. 
This method requires information about the positions of all particles, 
the ages of stellar components, gas properties as well as chemical 
composition. 

First, we constructed 2D regular Cartesian grids using the positional 
information of the particles. The spectral energy distribution (SED) is 
assigned to all stellar particles 
(both disk and bulge) following \citet{BC03}. Since our aim is to study the time 
evolution of merging galaxies at low redshift, potentially displaying resolved 
features, we assume a solar metallicity and 2 Gyr-old stellar components 
following \citet{Gallazzi05}. It is worth mentioning that dust has a critical role in 
diminishing the flux emitted by stars as the wavelength becomes shorter. 
Although the gas column density determines the strength of dust attenuation, 
the line-of-sight distribution 
of gas and stars should be considered for a more realistic approach. 
We use the empirical fitting formula from \citet{Bohlin78} and 
the dust extinction curve provided by \citet{Calzetti00} 
to calculate the amount of dust extinction from hydrogen column density. 
In their studies, the dust-to-gas ratios convolved with metal abundances 
are similar to those of Milky Way \citep{Alton98,Davies99}.

To construct a mock image, we determined the 
field-of-view (FOV), distance to a galaxy, and plate scale. We assume a 
luminosity distance, ${\rm d_L} =$ 100 Mpc, which represents a typical 
distance to nearby galaxies in the widely-used 
large-scale surveys such as the SDSS. Also, we assume a plate scale of 0.5 
arcsecond per pixel which is comparable to the SDSS plate scale ($\sim 0.4$ 
arcsecond per pixel), and a FOV of 6.8 arcmin which can physically cover the 
merging area. In this case, 0.5 arcsecond per pixel and 6.8 arcmin correspond 
to 0.24 kpc and 200 kpc, respectively. In the following sections, we analyze 
the observational properties of galaxy mergers using the SDSS filter system.

\subsection{Convergence tests}

Before presenting  our main results, it is necessary to conduct a series of 
convergence tests to estimate their robustness.

\subsubsection{Number of particles}

\begin{figure}
\centering
\includegraphics[width=0.5\textwidth]{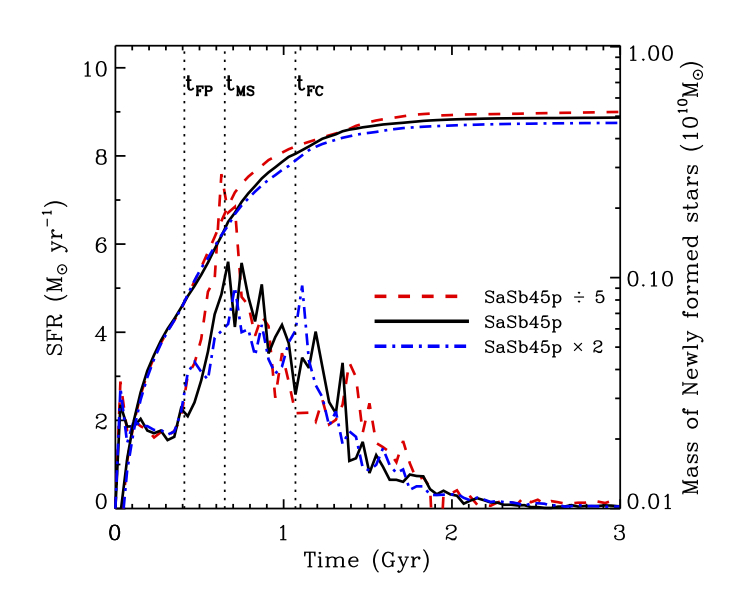}
\caption{Star formation rate and cumulative amount of newly-formed stars 
for various resolutions of the SaSb45p merger: the SaSb45p merger
(black solid line), five times fewer particles than the SaSb45p merger 
(red dashed line), and two times more particles than the SaSb45p merger 
(blue dashed dotted line).
}
\label{resolution}
\end{figure}

First,  we check to what extent the results are affected by the mass resolution. 
For this purpose, we ran the SaSb45p merger three times while varying the 
total number of particles used to model each galaxy/halo component, decreasing 
the number by a factor of five and increasing by a factor two. We 
changed the softening lengths to be inversely proportional to the number of 
particles, $N$. We used $N^{\rm -1/3}$ relation in this study 
\citep[see][]{Merritt96}.  
The gravitational softening lengths in the lower-resolution SaSb45p are 
$\epsilon_{\rm DM}$ = 0.17 kpc, $\epsilon_{\rm disk} = \epsilon_{\rm gas}$ = 
0.34 kpc, and $\epsilon_{\rm bulge}$ = 0.71 kpc, while those in the 
higher-resolution SaSb45p are $\epsilon_{\rm DM}$ = 0.079 kpc, 
$\epsilon_{\rm disk} = 
\epsilon_{\rm gas}$ = 0.16 kpc, and $\epsilon_{\rm bulge}$ = 0.079 kpc.
Figure \ref{resolution} shows the SFR and the time evolution of the amount of 
newly-formed stars for each simulation. Note that, despite slight 
differences, the SFRs all follow the same trend.  
Moreover, the difference between the final masses of new stars at 
$t=3$ Gyr is less than six percent, which suggests that our results are stable.

\begin{figure}
\centering
\includegraphics[width=0.45\textwidth]{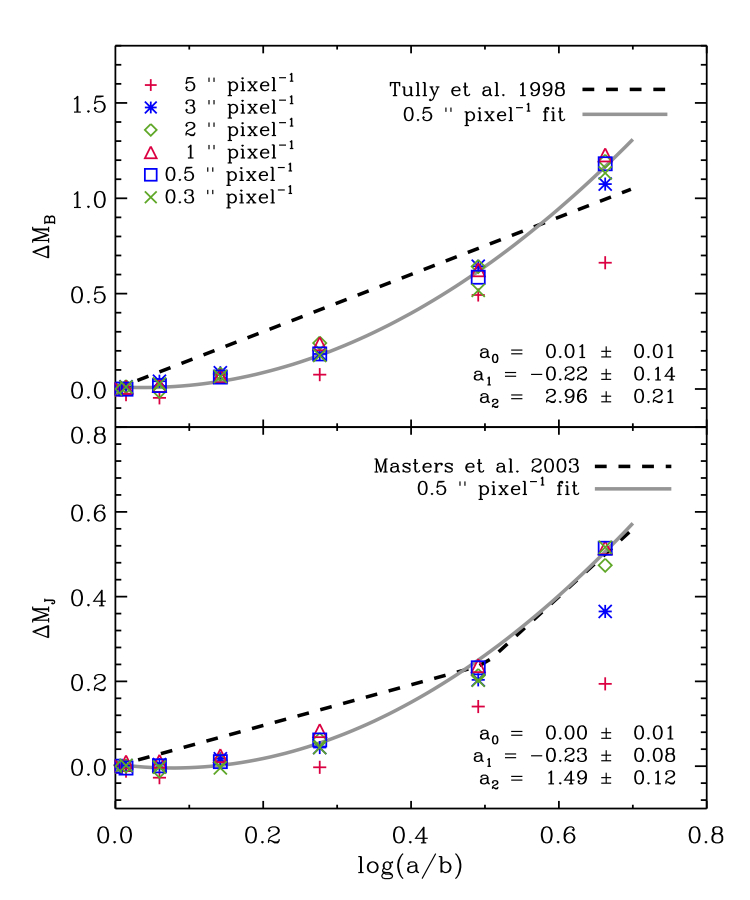}
\caption{Amount of dust attenuation in $B$ bands (top) and $J$ bands 
(bottom) with respect to the ratio between the major (a) and minor axis (b) of  
the Sb model galaxy. Each symbol represents the attenuated magnitude in 
each model galaxy with a specific plate grid resolution. The red dashed line in 
the top panel is the fitting function obtained from \citet{Tully98}, and the blue 
dashed line in the bottom panel is the bilinear fit obtained from 
\citep{Masters03}.}
\label{BJtest}
\end{figure}

\subsubsection{Plate scale}
It is also important to determine the optimal resolution at which we should 
perform the ray-tracing for the best accuracy in our predictions. 
Increased resolution requires more memory space and 
a longer computation time. In addition, finding a reasonable 
grid scale is the key to success in ray-tracing since, in this study, 
the physical size of the gas particle is ignored while  
we are able to assume stellar components as point sources.
 
Several studies on the internal extinction of spiral galaxies have been 
performed in 
the past decades \citep{Giovanelli94,Tully98,Xilouris99,Masters03}. The 
amount of dust attenuation is quantified through modeling an isolated galaxy. 
Methods for this computation can be put into two categories: analytical 
approximations \citep{Byun94,Baes01,Tuffs04} and Monte 
Carlo calculations \citep{Bianchi96,Matthews01,Pierini04}. 
However, models presented in these previous studies do not include galactic 
substructures such as spiral arms and clumps of gas  which are expected to 
affect the total amount of dust attenuation \citep{Corradi96,Rocha08}. 
Therefore, taking into account the presence of 
these substructures in the ray-tracing would be a necessary step toward realism.

We estimated the amount of dust attenuation 
using isolated late-type galaxies where the dependences 
of inclinations of galaxies are clear \citep[see e.g.][]{Byun94}. 
Figure~\ref{BJtest} shows the 
amount of dust extinction in the $B$ and $J$ bands in the 
Sb model galaxy as a function of inclination and grid resolution.
In particular, we compare the extinction slopes for six different plate scales
 as a function of axial ratio assuming the same luminosity distance. 
 
First, we quantified the amount of extinction in the $B$ band with respect to the 
grid resolution and the ratio between the major (a) and minor axes (b) of the Sb 
model galaxy. We found that, apart from the grid with the lowest resolution 
(5 arcsecond per pixel), the other grid resolutions (0.3 to 
3 arcsecond per pixel) led to similar variations in dust attenuation in 
the $B$ bands, 
$\Delta M_{\rm B}$ with respect to the axial ratio log$(a/b)$. When 
compared to the results from \citet{Tully98}, who measured $B$ band 
extinctions of late-type galaxies with various inclinations, we found that 
our choice of a grid resolution of 0.5 arcsecond per pixel 
agreed well with observations. Similar trends were found in 
the $J$ band. \citet{Masters03} argued that dust  
attenuation is not a linear function of the axial ratio, but is bilinear or has a 
quadratic form. \citet{Masters03} also suggest a bilinear equation of dust 
attenuation as a function of inclination compared to the simple linear 
formulation in \citet{Tully98}.

 Figure \ref{BJtest} shows that high grid resolutions lead to a  
nice agreement  with the observations. Low grid resolutions poorly resolve 
the structures which result in deviation from the fitting formula 
suggested by the observations. In the following, we then computed 
fitting functions by adopting grids of 0.5 arcseconds per pixel. We fit the 
$B$ and $J$ band extinction with a quadratic curve defined as follows:

\begin{equation}
{\Delta M_{\rm X}} =  a_0 + a_1 \log (a / b) + a_2 [\log (a / b)]^2
 \label{Jband}
\end{equation}

\smallskip
\noindent
where $a_0 = 0.01 \pm 0.01$, $a_1 = -0.22 \pm 0.14$, and $a_2 = 2.96 
\pm 0.21$ for the $B$ band, $a_0 = 0.00 \pm 0.01$, $a_1 = -0.23 \pm 0.08$, 
$a_2 = 1.49 \pm 0.12$ for the $J$ band, 
and $a/b$ is the axial ratio between the semimajor 
axis and the semiminor axis of a galaxy.

\begin{figure}
\centering
\includegraphics[width=0.45\textwidth]{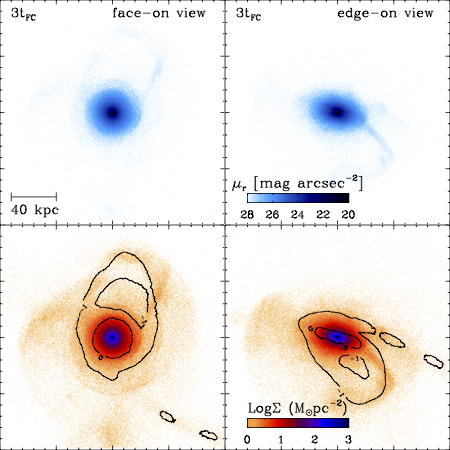}
\caption{
Mock image and surface column density of SbSb45h 
merger at three times the final coalescence time ($t_{\rm FC}$~=~1.57 Gyr).
Top panels show mock images viewed 
perpendicular to (face-on) and parallel to (edge-on)  
orbital plane of a companion galaxy in left-hand column and in 
right-hand column, respectively.
Color scheme is surface brightness in the SDSS $r$ 
band assuming luminosity distance, ${\rm d_L}$ = 100 Mpc 
and observational limit, $\mu_{\rm limit}$ =  28 mag arcsec$^{-2}$. 
Bottom panels show surface column density of stellar and gaseous mass 
viewed perpendicular to (face-on) 
and parallel to (edge-on) orbital plane of companion galaxy 
in left-hand column and in right-hand column, respectively.
Color scheme represents column density of stellar mass. 
Contour is gaseous surface density, and its values 
are labeled in logarithmic scale. }
\label{mu_SD}
\end{figure}

\begin{figure*}
\centering
\includegraphics[width=1\textwidth]{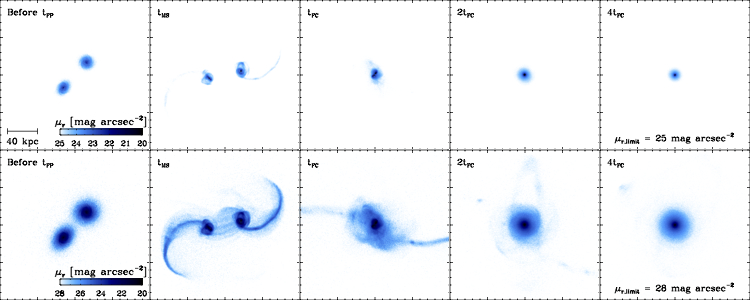}
\caption{Mock images of SbSb45h merger (mass ratio is 1:1)
viewed perpendicular to orbital 
plane of companion galaxy.
Color scheme is surface brightness in the SDSS $r$ 
band assuming luminosity 
distance, ${\rm d_L}$ = 100 Mpc.
Each row shows a merger process in time sequence: 
before the first passage time ($t_{\rm FP}$), 
the maximum separation time ($t_{\rm MS}$), and 
units of the final coalescence time ($t_{\rm FC}$), from left 
to right. Each column is plotted with different observation limits: 
$\mu_{\rm limit}$ =  25 mag arcsec$^{-2}$ (top) and 
$\mu_{\rm limit}$ =  28 mag arcsec$^{-2}$ (bottom).
Background color represents observation limit.
}
\label{mock_1:1}
\end{figure*}

\begin{figure*}
\centering
\includegraphics[width=1\textwidth]{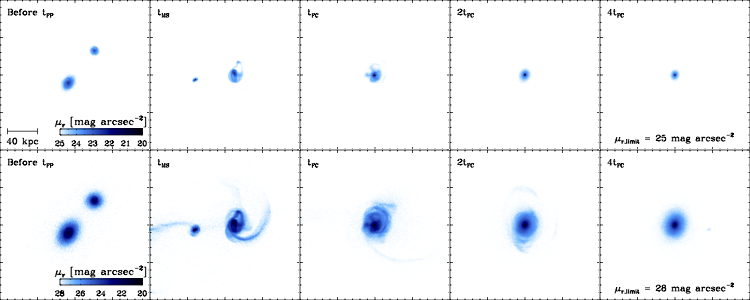}
\caption{Same as Figure \ref{mock_1:1}, but for 
${\rm SbSb_{3}45h}$ merger (mass ratio is 3:1)}.
\label{mock_3:1}
\end{figure*}

\begin{figure*}
\centering
\includegraphics[width=0.8\textwidth]{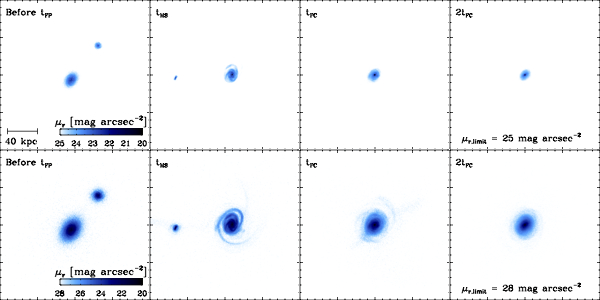}
\caption{Same as Figure \ref{mock_1:1}, but for 
${\rm SbSb_{6}45h}$ merger (mass ratio is 6:1).}
\label{mock_6:1}
\end{figure*}

\begin{figure*}
\centering
\includegraphics[width=1\textwidth]{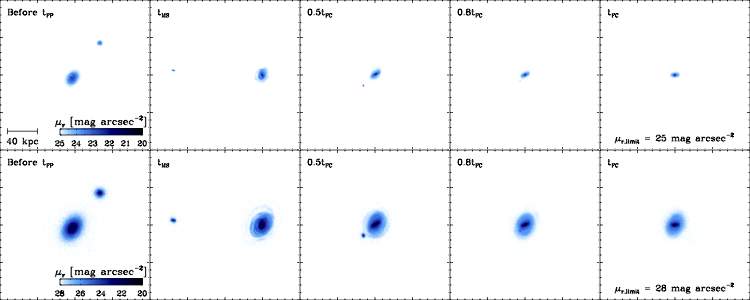}
\caption{Same as Figure \ref{mock_1:1}, but for 
${\rm SbSb_{10}45h}$ merger (mass ratio is 10:1).}
\label{mock_10:1}
\end{figure*}

\section{Results}
\label{results}

\subsection{Detectability of tidal features}

For collisionless particles, the dynamical timescale in the central 
region of a merger remnant is relatively short resulted from the violent 
mixture of the particles 
\citep{Mihos96,Springel00, Cox06b}. However, the outer region 
has a larger dynamical timescale which leads to long-lasting 
post-merger features such as shells, loops, and long tidal arms. 
For gas particles, they undergo shocks or dissipation of energy and angular 
momentum during the interaction. 
The gravitational torques resulted from perturbed structures induce 
gas to flow inwards \citep[see also][]{Negroponte83}. 
The gas inflows end up with the formation of a rotating rings or blobs.
 
Figure \ref{mu_SD} shows the mock images in the SDSS $r$ band 
(top panels) and stellar and gaseous column density (bottom panels) 
of the SbSb45h 
merger at three times the final coalescence time ($t_{\rm FC}$~=~ 1.57 Gyr). 
Although a few giga years elapsed after the final coalescence,  
post-merger features are clearly visible in the column density.
Substructures of merging galaxies are stretched out along 
with the orbital plane forming characteristic post-merger features 
such as shells and loops. 
Although the position and shape of the features are 
dependent upon a viewing angle,
they are clearly identifiable.

However, some of the post-merger features are 
invisible in mock images (see Figure \ref{mu_SD}, top panels). 
This is mainly because the features 
have low surface column density. 
In our simulations, stellar surface density lower than 
$\sim$ 2 ${\rm M_{\odot}~pc^{-2}}$ 
exceeds the surface brightness limit of $\mu_{\rm limit}$ =  
28 mag arcsec$^{-2}$.

\subsection{Merger-feature time of equal-mass mergers}

Figure~\ref{mock_1:1} shows the mock images 
of the SbSb45h merger at different characteristic times and according
to the observational limits of the SDSS $r$ band. 
The images using the SDSS $\mu_{\rm limit}$ reveal only the central 
part of each galaxy, resulting in the merger remnant reaching a spheroidal 
shape in a relatively short period of time. For example, the faint features 
of the SbSb45h merger become invisible already at 
twice the final coalescence time ($t_{\rm FC}~\sim$ 
1.57 Gyr) under the SDSS conditions. However, the mock 
images with a deeper $\mu_{\rm limit}$ 
indicate that the merger remnant exhibits extended faint structures for more 
than two times the final coalescence time. This confirms the fact that 
measuring the merger-feature time 
of merger remnants strongly depends on  $\mu_{\rm limit}$.

To compare two merger-feature times, we have estimated the ratio 
of the merger-feature time to the final coalescence time, 
${t_{\rm MF} / t_{\rm FC}}$ (see Table~\ref{mu28_time}).
Note that in contrast to 
the merger-feature time based on $\mu_{\rm limit}$ = 25 mag arcsec$^{-2}$, 
$t_{\rm MF,25}$, 
not all merger-feature times based on $\mu_{\rm limit}$ = 28 mag arcsec$^{-2}$, 
$t_{\rm MF,28}$ could be
estimated since faint features survive 
through the end of the simulation. Table~\ref{mu28_time} shows that 
${t_{\rm MF,25} / t_{\rm FC}}$ is always lower than three. The faint features in all  
merger simulations disappear by 2.16 $\pm$ 0.80 times the final 
coalescence time, or 1.38 $\pm$ 0.88 Gyr after the final coalescence, 
which is comparable to the finding of \citet{Lotz08}. However, 
${t_{\rm MF,28} / t_{\rm FC}}$ is approximately twice greater than 
${t_{\rm MF,25} / t_{\rm FC}}$ and exceeds three for all mergers. 

We also study how merger scenarios affect ${t_{\rm MF} / t_{\rm FC}}$.
The relaxation timescale of a merger remnant depends not only  
on orbital configurations \citep{Hernandez04} but also on the density of 
the merger remnant \citep{Conselice06}. 
Therefore, the relaxation timescales of the inner and outer regions of 
merger remnants are different, and the timescales are related to 
${t_{\rm MF,25} / t_{\rm FC}}$ or ${t_{\rm MF,28} / t_{\rm FC}}$. 
First, the prograde-prograde merger 
takes longer to have its substructures relaxed and 
thus to hide post-merger features 
than the retrograde-prograde merger does. 
For example, one can see in 
Table \ref{mu28_time} that the SbSb0p merger 
shows a larger value of ${t_{\rm MF,28} / t_{\rm FC}}$ than
the SbSb180p merger does, that is, 
6.22 $\pm$ 0.62 vs. 5.25 $\pm$ 0.41, respectively.
\footnote{
Much of our interpretation is based on the $\mu_{\rm limit} =$ 28 
mag arcsec$^{-2}$ cases, because $\mu_{\rm limit} =$ 25 mag arcsec$^{-2}$ 
cases usually reveal only the central regions of mergers which 
are insufficient to illustrate the whole merger effects.}
Second, mergers having a smaller orbital angular momentum 
(SbSb45r and SbSb45e mergers) 
tend to have shorter merging times ($t_{\rm FC}$, Column 4 in Table \ref{mu28_time}) 
and longer merger-feature times ($t_{\rm MF}$, Column 6 in Table \ref{mu28_time}) 
and hence resulting in larger values of ${t_{\rm MF} / t_{\rm FC}}$ 
(Column 8 in Table \ref{mu28_time}) 
than the larger-orbital-angular-momentum mergers (SbSb45p+ and SbSb45h).
We can perhaps interpret this as follows. Mergers with low orbital angular momentum 
(for example, direct head-on collisions) causes a rapid relaxation in the central 
region of the merger remnant. This makes the central region virtually act like a 
point source. The outer region of the merger remnant feels more loose and 
displays merger features for a longer period of time as a result. The essence 
in this interpretation is to realize that the merger timescale is determined by the 
structure of the central parts of the merger remnant whereas the merger feature 
timescale for deep imaging conditions is derived from the outskirts. 
Lastly, the bulge-to-total mass ratio (B/T) affects the 
relaxation timescales of at least the inner regions of the merger remnants. 
For this reason, all values of ${t_{\rm MF,25} / t_{\rm FC}}$ for the 
SaSb mergers are 
smaller than for the SbSb mergers.

\begin{figure}
\centering
\includegraphics[width=0.45\textwidth]{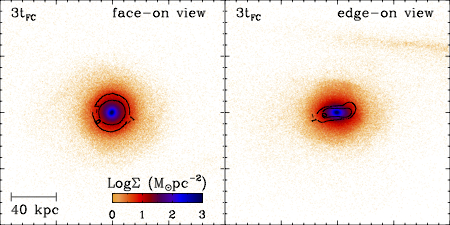}
\caption{
Surface column density of SbSb45hC  
merger at three times the final coalescence time ($t_{\rm FC}$~=~1.57 Gyr)
viewed perpendicular to (face-on) 
and parallel to (edge-on) orbital plane of a companion galaxy 
in left-hand column and in right-hand column, respectively.
Color scheme represents column density of stellar mass. 
Contour is gaseous surface density, and its values 
are labeled in logarithmic scale.}
\label{SD}
\end{figure}

\begin{figure}
\centering
\includegraphics[width=0.4\textwidth]{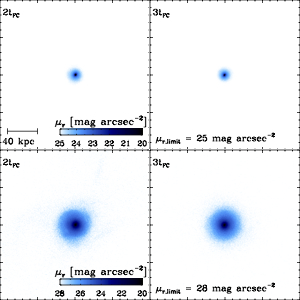}
\caption{Same as Figure \ref{mock_1:1}, but for 
SbSb45hC merger. Note that the evolution of the SbSb45hC merger 
before the final coalescence time is as the same as the SbSb45h merger.}
\label{mu_limits4}
\end{figure}

\begin{table*}
\begin{center}
\caption{Characteristic times and merger-feature time to final 
coalescence time ratio}
\begin{tabular}{lccccccc}
\hline\hline

Simulation &
${t_{\rm FP}}$ (Gyr)$^{\mathrm{a}}$ &
${t_{\rm MS}}$ (Gyr)$^{\mathrm{b}}$ &
${t_{\rm FC}}$ (Gyr)$^{\mathrm{c}}$&
${t_{\rm MF,25}}$ (Gyr)$^{\mathrm{d}}$ &
${t_{\rm MF,28}}$ (Gyr)$^{\mathrm{e}}$&
${t_{\rm MF,25} / t_{\rm FC}}^{\mathrm{f}}$&
${t_{\rm MF,28} / t_{\rm FC}}^{\mathrm{g}}$\\

\hline

\\
 \multicolumn{8}{c}{Equal-mass mergers in isolated environment} \\

SbSb0p  		&0.41	&0.71	&1.07		& 2.43 $\pm$ 0.34	& 6.65 $\pm$  0.66
	& 2.27 $\pm$ 0.32	& 6.22 $\pm$ 0.62	\\
SbSb45p  	&0.41	&0.65	&1.07			&2.31 $\pm$ 0.15	& 7.52 $\pm$  0.74
	& 2.16 $\pm$ 0.14	&7.03 $\pm$ 0.69 \\
SbSb90p  	&0.41	&0.76	&1.30			&3.37 $\pm$ 0.15	& 5.70 $\pm$ 0.45
	& 2.59 $\pm$ 0.11	& 4.38 $\pm$ 0.34 		\\
SbSb135p  	&0.41	&0.60	&1.13		&2.44 $\pm$ 0.12	& 6.13 $\pm$ 0.84
	& 2.16 $\pm$ 0.10	& 5.43 $\pm$ 0.84		\\
SbSb180p  	&0.41	&0.65	&1.03		&2.10 $\pm$ 0.24	& 5.41 $\pm$ 0.42
	& 2.04 $\pm$ 0.24	& 5.25 $\pm$ 0.41		\\ 

SaSb45p  	&0.41	&0.67	&1.09			&2.40 $\pm$ 0.33	& 5.63 $\pm$ 0.65
	& 2.20 $\pm$ 0.30	& 5.17 $\pm$ 0.59		\\
SaSb90p  	&0.41	&0.90	&1.41			&2.59 $\pm$ 0.22	& 4.82 $\pm$ 0.25
	& 1.83 $\pm$ 0.16	& 3.42 $\pm$ 0.18		\\
SaSb135p 	&0.41	&0.63	&1.08			&2.26 $\pm$ 0.20	& 4.99 $\pm$ 0.24
	& 2.09 $\pm$ 0.19 	& 4.62 $\pm$ 0.22		\\

SbSb45r	  	&0.38	&0.51	&0.73		&2.25 $\pm$ 0.26	& 7.01 $\pm$ 0.26
	& 3.08 $\pm$ 0.35	& 9.60 $\pm$ 0.35 \\
SbSb45p+  	&0.43	&1.00	&1.93		&3.04 $\pm$ 0.18	& 6.69 $\pm$ 0.32	
	& 1.58 $\pm$ 0.09	& 3.47 $\pm$ 0.17			\\
	
SbSb45e  	&0.49	&0.61	&0.84			&2.68 $\pm$ 0.32 	& $> 8.50^{\mathrm{h}}$
	& 3.19 $\pm$ 0.38	& $> 10.12^{\mathrm{h}}$\\
SbSb45h  	&0.35	&0.80	&1.57			&2.99 $\pm$ 0.20	& 5.60 $\pm$ 0.47
	& 1.90 $\pm$ 0.13	& 3.57 $\pm$ 0.30		\\
	
\hline

Mean	&0.41$\pm$ 0.03  &0.71 $\pm$ 0.14 &1.19 $\pm$ 0.33 
	&2.57 $\pm$ 0.82	&6.22 $\pm$ 1.73	& 2.16 $\pm$ 0.80	& 5.24 $\pm$ 1.46\\
\hline\\

 \multicolumn{8}{c}{Unequal-mass mergers in isolated environment} \\
 
${\rm SbSb_{3}45h}$ 	&0.44	&1.11	&2.13		&4.59 $\pm$ 0.17	& 7.26 $\pm$ 0.66
	& 2.15 $\pm$ 0.08			& 3.41 $\pm$ 0.31		\\ 

${\rm SbSb_{6}45h}$	&0.47	&1.48	&3.75		&5.80 $\pm$ 0.14	& 8.30 $\pm$ 0.39
	& 1.55 $\pm$ 0.04			& 2.21 $\pm$ 0.11		\\
	
${\rm SbSb_{10}45h}$	&0.48		&2.8		&12.9		
&-	&-  &- &-		\\
	
\hline

Mean$^{\mathrm{i}}$	&0.46$\pm$ 0.02  &1.30 $\pm$ 0.26 	&2.94 $\pm$ 1.15
	&5.19 $\pm$ 0.44	&7.78 $\pm$ 0.77	& 1.77 $\pm$ 0.15	& 2.65 $\pm$ 0.26\\
\hline\\



\multicolumn{8}{c}{Merger in a cluster potential } \\	
SbSb45hC 	&0.35	&0.80	&1.57		&2.81$\pm$ 0.17	&4.02 $\pm$ 0.17
	& 1.79 $\pm$ 0.11			& 2.56 $\pm$ 0.04		\\ 

\hline
\hline

\end{tabular}
\label{mu28_time}
\end{center}
\begin{list}{}{}
\item[$^{\mathrm{a}}$]First passage time.
\item[$^{\mathrm{b}}$]Maximum separation time.
\item[$^{\mathrm{c}}$]Final coalsecence time.
\item[$^{\mathrm{d}}$]Merger-feature time based on $\mu_{\rm r,limit}$ 
= 25 mag arcsec$^{-2}$.
\item[$^{\mathrm{e}}$]Merger-feature time based on $\mu_{\rm r,limit}$ 
= 28 mag arcsec$^{-2}$.
\item[$^{\mathrm{f}}$]Ratio between merger-feature time to final 
coalescence time based on 
$\mu_{\rm r,limit}$ = 25 mag arcsec$^{-2}$.
\item[$^{\mathrm{g}}$]Ratio between merger-feature time to 
final coalescence time timescales 
based on $\mu_{\rm r,limit}$ = 28 mag arcsec$^{-2}$.
\item[$^{\mathrm{h}}$]Disturbed structures are clearly visible until 
the end of simulation (6.5 Gyr).
\item[$^{\mathrm{i}}$]In this case, we consider only 
${\rm SbSb_{3}45h}$ and ${\rm SbSb_{6}45h}$ mergers.
\end{list}
\end{table*}

\subsection{Effect of mass ratio}

We run two additional mergers to qualitatively discuss the 
effect of mass ratio on the merger-feature time. 
Figure \ref{mock_3:1} and \ref{mock_6:1} show the mock images of 
the ${\rm SbSb_{3}45h}$ and the ${\rm SbSb_{6}45h}$ mergers, 
respectively. Their evolution is shown at different characteristic 
times and according to the observational limits of the SDSS $r$ band. 
For equal-mass mergers, two long and symmetric tidal arms are 
well developed as shown in Figure \ref{mock_1:1}. 
Unequal-mass mergers, however, show asymmetric tidal structures 
\citep[see also][]{Barnes92,Springel05b,Lotz08}.
As the mass ratio between two galaxies becomes larger 
(see Figure \ref{mock_6:1}), 
tidal bridges are unseen and tidal arms are tightly wound.

We investigate the effect of mass ratio on the merger feature time.
The ratio of the merger-feature time to the final coalescence time 
has no clear difference from major mergers (defined as mass 
ratios are less than 3:1).
For the ${\rm SbSb_{3}45h}$ merger, both ${t_{\rm MF,25} / t_{\rm FC}}$ 
and ${t_{\rm MF,28} / t_{\rm FC}}$ are comparable to those of the 
SbSb45h merger (see Table ~\ref{mu28_time}). 
On the other hand, although the merger-feature times of 
the ${\rm SbSb_{6}45h}$ merger (i.e. ${t_{\rm MF,25}}$ and 
${t_{\rm MF,28}}$)  are greater than those of the SbSb45h merger, 
both ${t_{\rm MF,25} / t_{\rm FC}}$ and ${t_{\rm MF,28} / t_{\rm FC}}$
of the ${\rm SbSb_{6}45h}$ merger are less than 
those of the SbSb45h merger.
In essence, unequal-mass mergers show merger 
features longer mainly because they take longer to 
merge, compared to equal-mass mergers.

Figure \ref{mock_10:1} shows a reference case of 10:1 merger. 
It is interesting to note that this ``minor'' merger case does not exhibit 
merger features clearly even before $t_{\rm FC}$. We are tempted to interpret 
this to suggest that such minor mergers, while they may be much 
more abundant than major mergers, may be very difficult to detect 
even in deep imaging conditions. Conversely, most of the merge 
features in today's typical surveys may very well be results of 
relatively ``major'' mergers.  Confirmation of this conjecture 
requires a much more thorough modeling effort.

It is also hinted in Figures \ref{mock_1:1} -- \ref{mock_10:1} that 
long symmetric tidal tails and large loops are found more 
frequently in major mergers \citep[see e.g.][]{Wang12}. 
It seems as well that the degree of symmetry scales to 
the mass ratio between the merging galaxies. It again 
requires many more runs filling a possible parameter 
space to confirm the statement. But if it is confirmed, it 
would imply that such features can be used to find 
and characterize major merger remnants.

\subsection{Effect of tidal force}

We also qualitatively examined the effect of tidal force on 
the merger-feature time.
Figure ~\ref{SD} shows the stellar and gaseous column density 
of the SbSb45hC merger at three times the final coalescence time 
($t_{\rm FC}$~=~1.57 Gyr). 
Stars and gas in low density region of the SbSb45hC merger are 
stripped off having undergone tidal force of a cluster potential.  
As a result, the stars and gas are concentrated in the central region 
unlike in isolated environment (see Figure \ref{mu_SD}).

Figure ~\ref{mu_limits4} shows the mock images of the SbSb45hC 
merger at twice and three times the final coalescence time and according 
to the observational limits of the SDSS r band. 
We found that the merger-feature time of $\mu_{\rm limit}$ =  25 mag 
arcsec$^{-2}$ for the SbSb45hC merger (${t_{\rm MF,25}}=2.81\pm0.17$) 
is comparable to the SbSb45h merger 
(${t_{\rm MF,25}}=2.99\pm0.20$).
Shallow mock images of $\mu_{\rm limit}$ =  25 mag arcsec$^{-2}$ mainly 
reveal the central region of each merger remnant which is 
less affected by tidal force because of its high density.

On the other hand, the effect of tidal force on post-merger features 
are better examined in deeper images. The merger-feature time of the 
SbSb45hC merger based on $\mu_{\rm r,limit}$ = 28 mag arcsec$^{-2}$ 
(${t_{\rm MF,28}}=2.56\pm0.04$) 
is $\sim$ 30\% smaller than 
the SbSb45h merger (${t_{\rm MF,28}}=3.57\pm0.30$).

\subsection{An empirical test on the SDSS database}

To test the depth dependence of merger feature lifetime, 
we performed a simplistic test using the SDSS database. 
We used the images of galaxies that are present both in 
the standard SDSS DR7 database and in the SDSS Stripe82 
database \citep{Abazajian09}. The surface brightness limit of 
Stripe82 (27mag in r' band) is roughly 2 mag deeper than that 
of DR7 instead of 3 mag which was the difference we made 
calculations for, and mergers that can be found in SDSS 
galaxies are not necessarily equal-mass mergers. Thus this 
comparison is not exactly suited to test our prediction; but it is 
the easiest test and probably serves as a useful milestone at least. 
We performed visual inspection on a volume-limited sample of 
1,453 galaxies. Among early-type galaxies, 59$\pm$4 and 160$\pm$14 
galaxies were classified as post-merger galaxies in DR7 and Stripe82, 
respectively, where the errors are from multiple inspections. 
While it is trivial that merger features are more frequent in deeper 
images, we are encouraged by the fact that roughly a factor of 3 
larger number of merger-feature galaxies are found in Stripe82 
as suggested by the simulations.

\section{Discussion}
\label{discussion}
In this paper, we investigate the merger-feature time 
of equal-mass disk mergers 
using N-body/hydrodynamic simulations. 
To cover a realistic range of merger scenarios, we consider 
different orbit types, host galaxy inclinations, 
and the morphological properties of the two main galaxies. 
We ran additional merger simulations considering different 
mass ratios and putting the mergers in a cluster halo 
environment to help understand the features of equal-mass 
mergers.

The post-merger signatures of nearly all samples 
in isolation survived for more than three times the 
final coalescence time when we see galaxies with a deep 
surface brightness limit of 28 mag arcsec$^{-2}$. 
The lifetime of merger features is shorter by 30\% 
when mergers happen in a large cluster environment. 
So, for random environments, the merger feature lifetime 
can be said to be $\sim$ 3.5 times the final coalescence time.
This has an obvious impact on the galaxy morphology 
classification, especially on post-merger remnants. 
\citet{Sheen12} recently found from their deep imaging 
($\mu_{\rm limit} \sim$ 28 mag arcsec$^{-2}$) campaign 
that a large fraction ($\sim$ 40\%) of bulge-dominant 
galaxies in massive clusters show major post-merger 
features. Using a semi-analytic model of \citet{Lee13}, 
\citet{Yi13} interpreted it as a result of merger 
relics from the merging events from the past halo 
environments by adopting the merger-feature time 
estimates from the preliminary result of our study. 

Even though we found and discussed noticeable 
differences in merger-feature time between a variety 
of proposed merger scenarios, our sample is still limited. 
For example, although equal-mass mergers are 
influential phenomena that clearly exhibit transitions 
of galaxy property, they are rare cases in the universe. 
Minor mergers are expected to dominate
galaxy evolution over major mergers because they 
occur more frequently \citep{Kaviraj09}.
However, our reference models suggest hints to that 
tidal features induced by relatively minor mergers 
(e.g., $m2/m1 \lesssim 1/6$) 
might be proved to be difficult to detect in deep images.

Moreover, mergers of gas-rich galaxies are 
expected to be frequent at high redshifts. Without 
considering gas-rich mergers, we cannot see the 
full picture of galaxy evolution. Furthermore, 
gas-rich mergers should accompany 
one of the most important effects on galaxy 
growth, namely AGN feedback. Recent studies 
have shown that super massive black holes (SMBHs) 
in the centers of galaxies are connected to the 
growth of galaxies. The velocity dispersion of the 
bulge component is proportional to the BH mass 
of the galaxy \citep{Tremaine02}. This coevolution 
of galaxies and SMBHs has been suggested by recent 
studies. For examples, the correlation between star 
formation and BH accretion \citep{Zheng09}, 
quasar activity of merging galaxies 
\citep{Sanders88,Hopkins08,Li07,Sijacki09}, and the 
quenching of star formation due to AGN feedback 
\citep{Springel05b,DiMatteo05,Schawinski06, Schawinski07a,
Schawinski07b, Sijacki07, Dubois10,Teyssier11,Dubois13}. 
For a more realistic 
and comprehensive study on merging galaxies, 
we will include the effects of AGNs in our future 
studies. However, we think 
that the inclusion of AGN feedback would not affect 
our analysis on post-merger features much because most of the 
star formation induced by merger is concentrated in the 
central regions of the merger remnant while post-merger 
features are mostly made up of older pre-existing stars 
in our simulations \citep[see also][]{Peirani10}.

We performed most of 
numerical simulations assuming an isolated 
environment. Galaxy interactions in reality occur in 
various environments including galaxy clusters. 
In addition to galaxy mergers, there are physical 
mechanisms affecting the evolution of galaxies in clusters: 
ram-pressure stripping \citep{Gunn72}, 
high-speed galaxy encounters 
\citep[galaxy ``harassment'',][]{Moore96}, 
tidal stripping by a cluster halo potential 
\citep{Byrd90,Valluri93}, and thermal evaporation 
of cold gas inside galaxies \citep{Cowie77}. 
Among these mechanisms,  tidal stripping 
plays a particularly important role in shaping merger 
remnants. 
Although we confirmed that 
the tidal structures of merger 
remnants are stripped off rapidly in a cluster potential 
\citep[e.g.][]{Mihos04}, we have not checked the 
effect of orbital parameters of merger 
remnants falling into galaxy clusters yet.
Therefore, the quantification of ${t_{\rm PM} / t_{\rm FC}}$ 
in various cluster environments 
should be investigated in future studies.

Ram pressure stripping, and other stripping processes, 
affect the gas content in galaxies in deep potential wells 
which will in turn affect any gas-dependent properties of 
galaxies, e.g. star formation. However, it would not 
significantly affect our main predictions on merger-feature 
times. This is because, as our simulations suggest, most 
of the faint merger features that are easily detected are 
composed of pre-existing old stars (which behave following 
dissipationless processes and angular momentum conservation) 
rather than newly formed stars \citep[see for instance Figure 4 of][]{Peirani10}.
Moreover, we do not expect that Ram pressure stripping 
has a significant impact on removing gas in our model 
galaxies (M $\sim$ 1.7 $\times 10^{11} {\rm M}_{\odot}$ for equal-mass mergers) 
because it is thought to be more efficient for lower mass objects, 
i.e., M < 6 $\times 10^{10} {\rm M}_{\odot}$ \citep{Nickerson11}. 
We admit that more robust 
calculations must include all important physical processes, 
such as Ram pressure stripping, in the end. But for the 
focus of this paper at the moment, current calculations 
should be sufficiently elaborate.

\begin{acknowledgements}
We thank Volker Spingel for making the Gadget 
code available to us. SKY acknowledges support 
from National Research Foundation of Korea 
(NRF-2009-0078756; NRF- 2010-0029391) and 
DRC Grant of Korea Research Council of Fundamental 
Science and Technology (FY 2012). 
Numerical simulation was performed using the 
KISTI supercomputer under the program of 
KSC-2012-C2-11 and KSC-2012-C3-10
and the KASI supercomputer. Much of 
this manuscript was written during the visit of SKY 
to University of Nottingham and University of Oxford 
under the general support by LG Yon-Am Foundation. 
we would like to thank the referee for his/her comments 
that have improved the quality of the original manuscript.
\end{acknowledgements}

\bibliographystyle{aa} 

\end{document}